\documentclass[10pt]{article}
\usepackage[OE]{express}

\begin{document}
\title{Radiative cooling by tailoring surfaces with microstructures}

\author{Armande Herv\'{e},\authormark{1,*} J\'{e}r\'{e}mie Drevillon,\authormark{1} Youn\`{e}s Ezzahri,\authormark{1} and Karl Joulain \authormark{1}}

\address{\authormark{1}Institut Pprime, CNRS, Universit\'{e} de Poitiers, ISAE-ENSMA,F-86962 Futuroscope Chasseneuil, France}

\email{\authormark{*}armande.herve@univ-poitiers.fr} 



\begin{abstract}
We propose in this article a method to generate radiative coolers which are reflective in the solar spectrum and emissive in the transparency window of the atmosphere (8-13 $\mu$m). We choose an approach combining thermal control capacity of gratings and multi-layers. We use optimized BN, SiC and SiO$_{2}$ gratings, which have emissivity peak in the transparency window. We place under these gratings a metal/dielectric multi-layer structure to obtain a near perfect reflectivity in the solar spectrum and to enhance the emissivity in the transparency window. The optimized structures produce a good radiative cooling power density up to 80 W.m$^{-2}$.
\end{abstract}


\section{Introduction}
The combined improving demand in housing comfort and temperature rise due to global warming, has increased energy consumption in air conditioning. This strongly impacts the environment so that limiting cooling energy expenses has become a real issue. One possible option to fulfill this goal is through material design for a radiative cooling which in principle is very simple: evacuate the heat directly into space.
Indeed, it is well known that the earth's atmosphere has a transparency window for electromagnetic waves between 8 and 13 $\mu$m. This transparency window coincides with thermal radiation wavelengths at typical ambient temperatures. Using this phenomenon, a body can be cooled just because its heat is radiated into cold outer space: this is the passive radiative cooling process. 

Radiative cooling is a common phenomenon at the earth's surface and it can be illustrated by several natural processes. For example dew or frost formation observed in the morning after long winter nights is a well-known manifestation of such a phenomenon. Nighttime radiative cooling systems have been extensively studied \cite{catalanotti_radiative_1975,bartoli_nocturnal_1977,harrison_radiative_1978, granqvist_surfaces_1980, granqvist_radiative_1981,granqvist_radiative_1982,berdahl_radiative_1984,berdahl_thermal_1983, orel_radiative_1993,smith_amplified_2009,gentle_radiative_2010, suryawanshi_radiative_2009}. Many selective emitters in the spectral window were exploited : cheap plastic \cite{catalanotti_radiative_1975, bartoli_nocturnal_1977}, pigmented paints \cite{harrison_radiative_1978, orel_radiative_1993}, SiO and Si$_{3}$N$_{4}$ films \cite{granqvist_radiative_1981,granqvist_surfaces_1980, granqvist_radiative_1982} and more recently composite materials \cite{gentle_radiative_2010, suryawanshi_radiative_2009}. Other authors have compared different kinds of materials and their properties \cite{berdahl_radiative_1984, berdahl_thermal_1983, smith_amplified_2009}.
\bigskip

However, the cooling demand is much more important at daytime, and if radiative cooling occurs naturally at night in the absence of daylight, it becomes complicated at daytime because of the heating by the sun that influences the radiative cooler.
To produce daytime radiative cooling, a strong emission in the transparency window and a quasi-total reflection (90 \% or more \cite{rephaeli_ultrabroadband_2013}) in the solar spectrum are required. However, it is difficult to achieve simultaneously these two properties.
Some previous studies have tried to design daytime radiative coolers \cite{nilsson_solar_1992, nilsson_radiative_1995} but without much success due to the fact that the reflection rate was not sufficiently important to prevent overheating of the structure.
In 2013, Rephaeli et al. \cite{rephaeli_ultrabroadband_2013} experimentally proved for the first time the concept of daytime passive radiative cooling with a new structure based on a photonic crystal. The structure used as a cooler made of two parts: a 2D photonic crystal using phonon-polariton mode to obtain maximum emissivity in the infrared atmospheric transparency window and a 1D photonic crystal to reflect radiation in the solar spectrum. In 2014, it was shown that a metamaterial could also act as a daytime radiative cooler \cite{hossain_metamaterial_2015}. The device was composed of an anisotropic and conical-shaped metamaterial structure, for a better polarization insensitivity. It presented a large selective infrared emission on the entire atmospheric transparency window (8-13 $\mu$m). These two results demonstrated that it is possible to realize energy-efficient radiative cooling devices but the proposed structures are quite complex to fabricate and they do not seem suitable for a possible mass production. At the end of 2014, a system composed of a nanophotonic solar reflector and a thermal emitter was designed \cite{raman_passive_2014}. When exposed to a direct solar irradiance of 850 W.m$^{-2}$ on a rooftop, the structure has reached a cooling power density of 40 W.m$^{-2}$ at ambient temperature. In 2017, an ingenious type of device was proposed for daytime radiative cooling \cite{zhai_scalable-manufactured_2017}, which manufacturing seems easy and economical. The structure is a randomized glass-polymer hybrid metamaterial. It is transparent polymer with silicon dioxide microspheres inside disposed at random. The emission in the atmospheric transparent window comes from the microspheres employing phonon-enhanced Fr\"ohlich resonances while the reflection of solar irradiance is performed with a thick silver coating. 
Note that many other structures have also been proposed either for solar cells cooling \cite{zhu_radiative_2014, zhu_radiative_2015, safi_improving_2015, wu_solar_2015, zhou_radiative_2016} or just for daytime radiative cooling \cite{huang_nanoparticle_2017, kecebas_passive_2017, kou_daytime_2017}.

\bigskip

In this work, our aim is to propose a new design, which combines the abilities of thermal emission control of gratings and multi-layer structures. To this end, we choose to associate thin films stack with surface gratings. Indeed, it is well known that one is able to control spectrally and directionally  the thermal emission by ruling a grating on the surface of a polar material, supporting surface phonon-polaritons, such as SiC (Silicon Carbide) or SiO$_{2}$ (Silicon Dioxide - $\alpha$-Quartz) \cite{greffet_coherent_2002, marquier_coherent_2004, dahan_space-variant_2005, arnold_coherent_2012}.
Coupling these gratings with other structures, such as multi-layers structure, which also have emission properties different from classical lambertian thermal sources \cite{drevillon_far_2011, nefzaoui_selective_2012}, is therefore a natural way to control and shape surface radiative properties and obtain structures with very particular emissive properties. It could indeed complete the grating emission into the atmospheric window.

In the next sections, we  first remind the principles of radiative cooling. Then, we present our structures, their properties and the calculation methods used to optimize and calculate their cooling power density. Finally, we present the final optimized structures and discuss their performances.

\section{Principles of radiative cooling}               

\begin{figure}[hbtp]
 \centering
 \includegraphics[width=8cm]{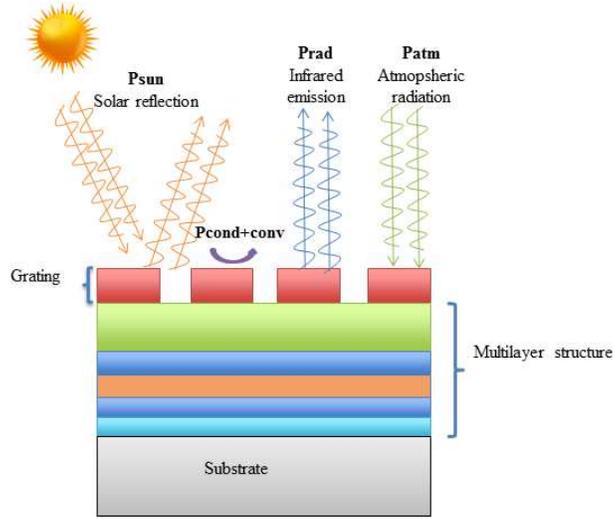}
 \caption{Schematics of a radiative cooler composed of multi-layer structure and grating overhead with some of the radiative and nonradiative processes.}
 \label{schema Pcool}
 \end{figure}

We define the radiative cooling power $P_{cool}$ \cite{raman_passive_2014} as follows (Fig.~\ref{schema Pcool}):
\begin{equation}
P_{cool} (T) = P_{rad} (T)-P_{atm} (T_{amb})-P_{sun}-P_{cond+conv} 
\end{equation}

In this equation, $P_{rad}$ is the power emitted by the structure, defined by equation (2).
\begin{equation}
P_{rad} (T) = A \int d\Omega cos(\theta)\int_{0}^\infty d\lambda I_{BB}(T,\lambda) \epsilon(\lambda,\Omega)  
\end{equation}
where $\int d\Omega=\int_0^{\pi/2} sin \theta d\theta \int_0^{2\pi}d\phi$ is the angular integral between $\theta=0$ and $\pi/2$ and between $\phi=0$ and $2\pi$ for a hemisphere, $I_{BB}(T,\lambda)$ is the black body specific intensity at temperature $T$, $\epsilon(\lambda,\Omega)$ is the spectral and angular emissivity of the structure and $A$ the surface of the device. 
Here, we need to be careful with the structure we study, because the emissivity of a plane surface like the multi-layer does not depend on the polarization and the azimuth angle, which is not the case for the grating, where emissivity can be different for both polarizations and also depend on the azimuthal angle.
So we take into account in the expression above the azimuthal angle $\phi$ in addition to the emission angle $\theta$ and the emissivity should also be a function of the wavelength $\lambda$ and space angle $\Omega$.  
However, we will show in the next parts, that we choose omnidirectionnal gratings, i.e. which have emissivity peaks for certain wavelengths for all angle of incidence. Marquier et al. \cite{marquier_anisotropic_2006,marquier_degree_2008} have shown that for omnidirectionnal grating sources, the azimuthal dependance is negligible. We have checked for our structures and it is effectively the case. So the only difference is the emissivity of the grating between TM and TE polarizations.
So we could use a simplified expression $\int d\Omega=\int2\pi* sin \theta d\theta$ of the angular integral between $\theta=0$ and $\pi/2$ for a hemisphere and use $\epsilon(\lambda,\theta)$ as the spectral and angular emissivity of the structure.
But to have a realistic result, we need to take for the structure the average emissivity between TM and TE polarizations. For the rest of the study and for sake of simplicity, the structure emissivity will only be presented for one of the polarization (TM), but the radiative cooling power will be calculated from the average emissivity between TM and TE polarizations.

\bigskip

$P_{atm}$ is the power from incident atmospheric radiation:
\begin{equation}
P_{atm} (T_{amb}) = A \int d\Omega cos(\theta)\int_{0}^\infty d\lambda I_{BB}(T_{amb},\lambda) \epsilon(\lambda,\theta) \epsilon_{atm}(\lambda,\theta) 
\end{equation}
The atmospheric emissivity is given by: $\epsilon_{atm}(\lambda,\theta)= 1-t(\lambda)^{1/cos(\theta)}$ where $t(\lambda)$ is the atmospheric transmittance in the zenith direction.

$P_{sun}$ is the absorbed power incoming from the sun:
\begin{equation}
P_{sun} = A \int_{0}^\infty d\lambda \epsilon(\lambda,\theta_{Sun}) I_{AM1.5} (\lambda)
\end{equation}
with the solar illumination represented by $I_{AM1.5}$, which correspond to the solar spectrum after it goes through 1.5 times the atmosphere thickness. We consider our structure facing the sun with the angle $\theta_{Sun}$, which is the sole angular dependence in the emissivity.

The conduction and convection transfer is given by a simple Newton law in $P_{cond+conv}$ (5) with a coefficient $h_{c}$ combining both heat transfer mechanisms.
\begin{equation}
P_{cond+conv} (T,T_{amb})= A h_{c} (T_{amb} -T)
\end{equation}

To produce radiative cooling, the resulting power $P_{cool}$ should be positive. 
When $P_{cool}(T)=0$, the temperature of the structure is corresponding to the steady state temperature $T_s$. This is an equilibrium temperature; if the external conditions do not change, the structure should stay at this temperature. So if $P_{cool} (T)>0$, the power excess represents the radiative cooling power. If the structure stays in the same external conditions, its temperature should naturally decrease.

To have a good radiative cooling power, an emissivity near 1 for most wavelengths in the [8-13]$\mu$m range and for most angles of incidence is really useful, because it increases significantly the power emitted by the optimized structure $P_{rad}$ without influencing the other power $P_{sun}$ and $P_{atm}$. These two powers tend to decrease $P_{cool}$, so it is interesting to keep them as low as possible. Other possibilities to improve the radiative cooling are:

- to keep a low solar absorption, the maximum allowed solar absorption is 10 percent in order to compensate it with the emission in the atmospheric window \cite{rephaeli_ultrabroadband_2013}. That is quite done in our structure with the association Ag - multi-layer;

- to reduce at most the convection and conduction around the structure by isolating it.

\section{Approach of problem}               
      
\begin{figure}[hbtp]
 \centering
 \includegraphics[width=7cm]{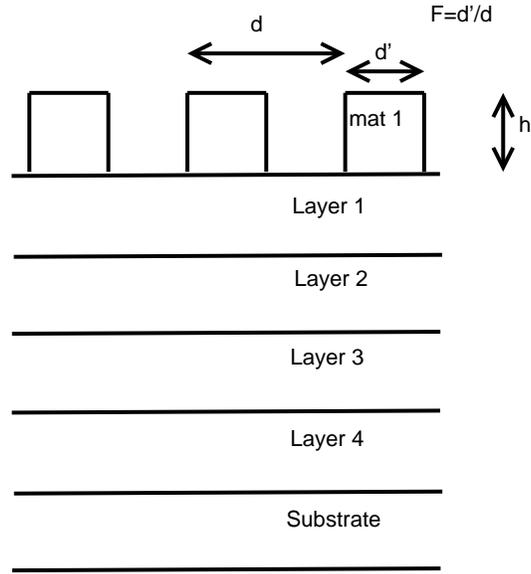}
 \caption{Schematic representation of the studied structure.}
 \label{schema}
 \end{figure}

The studied structure is constituted of three components: a metallic layer used as a substrate is placed under a multi-layer with a lamellar grating ruled on the top of it (Fig.~\ref{schema}). Note that it is the first time that a grating is used to conceive daytime radiative coolers. 
The aim is to optimize this structure in order to have a perfect reflexion in solar spectra and a perfect emission in the atmospheric window [8-13] $\mu$m. We could then have daytime radiative cooling and reduce our structure temperature below ambient temperature.
 
\subsection{Metallic layer}  
To generate a good cooling, it is essential to have a quasi-perfect reflection in the solar spectrum. We can achieve this with a metal layer, used as a substrate. A study of the optical properties of several metals was performed and silver showed up as the best candidate. So we use silver as the metallic layer for the rest of the study.

\subsection{Multi-layer structure contribution}  
The multi-layer above may be composed of materials with interesting optical properties ( high or low index, optically transparent, SPP (surface phonon-polaritons), ...). On the one hand, it should permit to increase the solar spectrum reflexion and on the other hand, it should contribute to the thermal emission in the earth atmosphere transparency window. In association with the grating, it can inhibit or increase the emission on a large spectra or create new emission peaks, especially if the multi-layer is made of materials supporting SPPs. Alternating transparent materials with absorbing materials can be used to create a resonant cavity\cite{drevillon_far_2011, nefzaoui_selective_2012}. At last, a layer of inert material could be inserted between two layers for fabrication issues.

\subsection{Grating contribution}
The grating is a new element in regards to daytime radiative cooling literature. It permits to enhance the emission between 8 and 13 $\mu$m.
The three parameters defining the grating are the period $d$, the depth $h$ and the filling factor $F$ (Fig.~\ref{schema}).
SiC, SiO$_{2}$ and BN (Boron Nitride) are the 3 materials that have been used for the grating. The reason is that these three materials are polar so that they support SPP, which are evanescent waves bordered in the near-field close to the interface. Note that these surface waves appear only in p-polarization (or transverse magnetic). When a grating is ruled at the source surface, it scatters the surface wave and couples it to a propagative wave in the far-field.

In principle, the 3 materials chosen for the grating are good candidates since they exhibit SPP at wavelengths located in the transparency window 8-13 $\mu$m. The grating should therefore scatter the thermally excited SPP to the far-field providing strong emission close to the resonance.

\subsection{Numerical method (RCWA with PSO)}
To calculate the radiative properties of our multi-layer with grating structure as a function of the wavelength $\lambda$ and the angle of incidence $\theta$ of the incident field, we used the Rigorous Coupled-Wave Analysis (RCWA) method. We used a code based on RCWA, RETICOLO, developed by J.P Hugonin and P. Lalanne from Institut d'optique. A short description and references to the RCWA method and RETICOLO could be found in \cite{herve_temperature_2016}.

For this study, we combined RETICOLO with a particle swarm optimization (PSO) algorithm. PSO was first presented by Kennedy and Eberhart \cite{kennedy_particle_1995}. It is built on a collaborative system with different particles, which allows to converge towards minimums. There are three main actions in a PSO algorithm : evaluate, compare and imitate. To converge towards the target, each particle decides its next move from its current speed, its best solution and the best solution of its neighbours (its informants)\cite{clerc_optimisation_2004}. 
Here, the target fixed for the PSO algoritm is the radiative properties of an ideal daytime radiative cooler which emits only between 8 and 13 $\mu$m and is perfectly reflective elsewhere for all angles of incidence.

We want to optimize the type of materials and the thickness for each layer and the three parameters of the grating, to obtain an emissivity as a function of the wavelength and the emission angle as close as possible of the ideal target defined above for both states of polarization of light. The research space is very large with several local minima and no unique solution. Furthermore, the optimization should deal with discrete variables, as the type of material for the layer. The PSO method is particularly suitable for this kind of optimization with discrete and continuous variables. Our team has already used this kind of algorithm before \cite{nefzaoui_selective_2012} to optimize multi-layer structures and further details could be found in \cite{nefzaoui_selective_2012}. For this study, we choose to optimize first structures with only 4,5 or 6 layers and to reduce the number of material types.

\section{Results and discussion}
\subsection{Results }
\begin{figure}[hbtp]
 \centering
 \includegraphics[width=7cm]{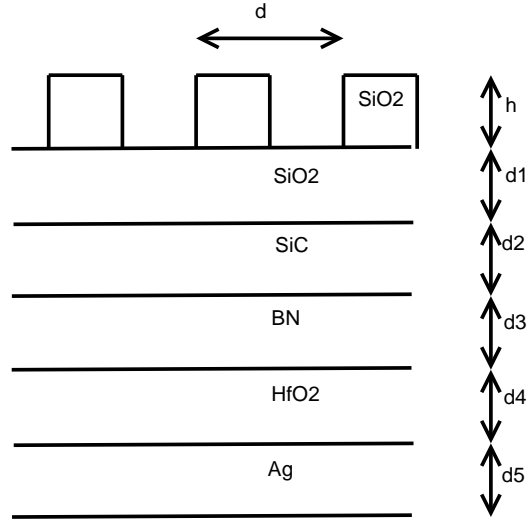}
 \caption{Optimized multi-layer structure with SiO$_{2}$ 1D grating.}
 \label{schema2}
 \end{figure}

We found several structures corresponding to our criteria, which produce a good radiative cooling, but we will examine thoroughly only one of them. This optimized structure combines a SiO$_{2}$ 1D grating with a multi-layer structure Ag/HfO$_2$/BN/SiC/BN/SiO$_2$ as seen in Fig.~\ref{schema2}. Its parameters are listed as structure 1 in Table~\ref{Table1}. We add the characteristics of structures 2 and 3, which also permit to have a good radiative cooling, but with a SiC and a BN grating respectively. We will not examine thoroughly these structures here, but it shows that with a good optimization, it is possible to have an important radiative cooling with the 3 types of materials for the grating. 

\begin{table}
\centering
\begin{tabular}{|p{0.5in}|p{0.28in}|p{0.27in}|p{0.27in}|p{0.39in}|p{0.39in}|p{0.39in}|p{0.39in}|p{0.39in}|}
\hline Structure & \multicolumn{3}{c|}{Grating $d$($\mu$m), $F$ et $h$($\mu$m)} & Layer 1 $d1$($\mu$m) &  Layer 2 $d2$($\mu$m) & Layer 3 $d3$($\mu$m) & Layer 4 $d4$($\mu$m) & Layer 5 $d5$($\mu$m)  \\
\hline  1 & \multicolumn{3}{c|}{SiO$_{2}$} & SiO$_{2}$ & SiC & BN & HfO$_{2}$ & Ag  \\
\cline{2-9}  & 3.37 & 0.39 & 0.736 & 0.93 & 0.725 & 0.999 & 0.08 & 0.11 \\
\hline 2 & \multicolumn{3}{c|}{SiC} & SiC & SiO$_{2}$ & SiC & BN & Ag  \\
\cline{2-9}  & 3.28 & 0.4 & 0.457 & 0.03 & 0.808 & 0.812 & 0.942 & 0.929 \\
\hline  3 & \multicolumn{3}{c|}{BN} & BN & SiC & SiO$_{2}$ & Ag & /  \\
\cline{2-9}  & 3.83 & 0.59 & 0.998 & 0.0711 & 0.686 & 1.0313 & 0.99 & / \\
\hline 
\end{tabular}
\caption{Parameters of our optimal multi-layer + grating structures for SiO$_2$, SiC and BN gratings.}
\label{Table1}
\end{table}

\begin{figure}[hbtp]
 \centering
 \includegraphics[width=8cm]{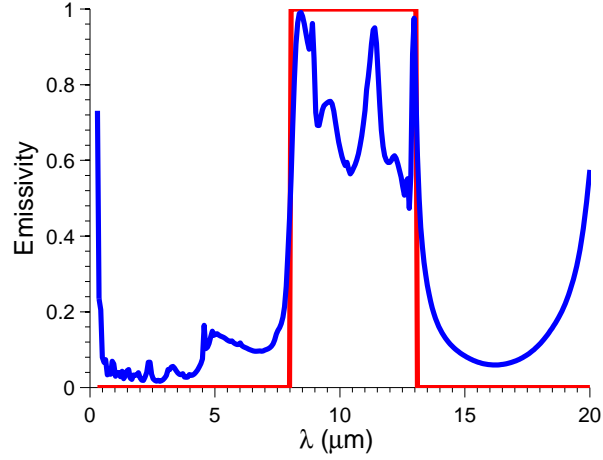}
 \caption{Emissivity at normal incidence in polarization TM of the optimized structure as a function of wavelength (blue line) compared to the ideal emissivity (red line).}
 \label{E_ld_SiO2_2}
 \end{figure}

\begin{figure}[hbtp]
 \centering
 \includegraphics[width=8cm]{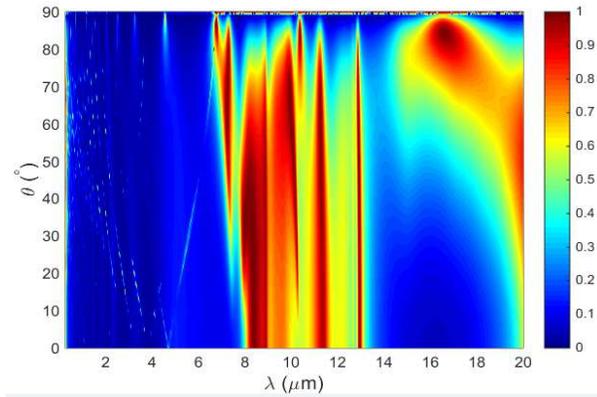}
 \caption{Emissivity in polarization TM of the optimized structure as a function of wavelength and angle of incidence.}
 \label{E_ld_teta_SiO2_2}
 \end{figure}

We plot in Fig.~\ref{E_ld_SiO2_2} the emissivity at normal incidence of our structure 1 as a function of wavelength compared to the ideal emissivity and in Fig.~\ref{E_ld_teta_SiO2_2} the emissivity as a function of wavelength and angle of incidence.
 
In Fig. ~\ref{E_ld_SiO2_2}, we see emissivity peaks close to 1 in the [8-13] $\mu$m wavelength range thus coming close to the ideal emissivity that we had fixed in the PSO. Note that this result is different from the one obtained with simple gratings. Indeed,  a SiO$_{2}$ 1D grating has a typical single emissivity peak located around 9 $\mu$m. The multi-layer structure completes very well these peaks by adding it own peaks in the whole atmosphere transparency zone. Some interactions between the grating and the multi-layer can modify further the emissivity in this area. 
On the other side, in the solar spectrum, the emission is reduced to a single peak for $\lambda$<0.4 $\mu$m, this peak has very little impact on $P_{sun}$ and then on $P_{cool}$. For $\lambda$>0.4 $\mu$m, the emission is near zero, so that the solar absorption by the structure is reduced and it contributes to a good radiative cooling. 
Mapping these results as a function of the wavelength and the angle of incidence (Fig.~\ref{E_ld_teta_SiO2_2}), we can see that the emissivity is nearly independent of the angle of incidence. Although the transmittance of the atmosphere decreases with the angle of incidence, as shown in the supplementary materials of \cite{chen_radiative_2016}, it is still high up to values close to 60 degrees. Therefore, an isotropic emission is more efficient for a good radiative cooling. 
\bigskip

\begin{figure}[hbtp]
 \centering
 \includegraphics[width=8cm]{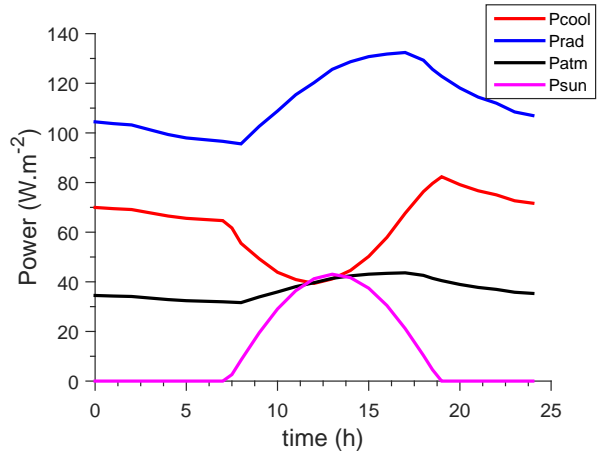}
 \caption{Radiative cooling power density and its contributions (with the average emissivity of the structure between TM and TE polarizations) as a function of time regarding to the temperature of a typical March day in Poitiers.}
 \label{Pcool_SiO2_2}
 \end{figure}

To evaluate the performance of our structure, we show in Fig.~\ref{Pcool_SiO2_2} the variation of the calculated radiative cooling power density (in red) during a clear day in our city Poitiers (located at the west of France with an oceanic and temperate climate). As an example, we took the ambient temperature of the clear day of 16 March 2017 in Poitiers. We want to clarify that in this example, we are not in the most favorable situation, because the temperature are not really high for the light illumination, so a better temperature/ light illumination ratio could produce more radiative cooling power density. But we choose this example to have an idea of the producible radiative cooling power density along a lambda day in our location, to help for example for further experimental testing. 
We suppose that the structure temperature is equal to the ambient temperature during the day. It permits on the one hand, to simulate a case where we are totally isolated from conduction and convection. On the other hand, it is similar to considering that the radiative cooling power produced by our structure is the power needed to make these two temperatures stay equal.
To calculate this radiative cooling power we took, as we have mentioned before, the average emissivity between TM and TE polarizations.
As we can neglect $P_{cond+conv}$, we can see in Fig.~\ref{Pcool_SiO2_2} the different contributions of $P_{cool}$ as defined in equation (1).

As could be expected, it is at night where we have the highest mean  $P_{cool}$ around 70 W.m$^{-2}$. The maximum value is reached at 7pm around sunset, when the temperature is still very high, but the sun contribution has almost disappeared. At noon, when the sun contribution is at its maximum, we still have a radiative cooling power density of 40 W.m$^{-2}$. Our performances are of the same order of magnitude as other daytime radiative coolers proposed recently \cite{rephaeli_ultrabroadband_2013, raman_passive_2014, kecebas_passive_2017}. Hence, a good radiative cooling could be obtained by using a relative simple structure combining a multi-layer and a 1D grating.
  
 \bigskip 
  
 Few interpretations about the emissivity of our optimized structure are given in the next part.

\subsection{Results interpretation}

Concerning the multi-layer structures, we have considered in a first approach only metals, HfO$_2$ and polar materials in their constitution.

We took a metallic layer as a substrate of our multi-layer structure in order to reduce the solar absorption. To reflect most of the solar radiation, silver seems to be the best metal with an affordable price. This reflection will be completed by the dielectric layer above.
The HfO$_{2}$ layer does not modify a lot the emission in the transparency window, but it helps to further reduce emission peaks in the solar spectrum in association with upper layers and it can serve as a buffer layer in order to facilitate the layer deposition.
  
The upper part is constituted of polar materials, which have the particularity to support SPP. The multi-layer can produce emissivity peaks different from the grating through resonance or cavity effect. It is also possible that these non radiative surface waves could be coupled through the grating to a propagative wave in the far-field. Another possibility is that the grating could act as an upper layer added to the multi-layer and modify or enhance some resonances already existing in the multi-layer. So the multi-layer structure can increase emissivity of some peaks and create others. However, unlike simple gratings, whose we can predict position and origin of peaks, it becomes difficult for a mix of multi-layers and gratings. That's why PSO is useful to optimize our structures with emission peaks only in the [8-13] $\mu$m range.

For the grating, the three types of materials (SiO$_{2}$, SiC and BN) can potentially produce good radiative cooling, because all emissivity peaks are located in the transparency atmospheric window [8-13] $\mu$m. It is also important to choose well the grating period. For these three polar materials, small period tends to produce monochromatic and isotropic emission peaks and for larger periods, the emission peaks are monochromatic and directional. Our optimized gratings have all small periods and produce consequently isotropic emission peaks \cite{marquier_coherent_2004, marquier_degree_2008}. The grating could also scatter thermally excited SPP from the multi-layer to the far-field adding therefore to the peaks produced solely by the grating and the multi-layer.

\begin{figure}[hbtp]
 \centering
 \includegraphics[width=8cm]{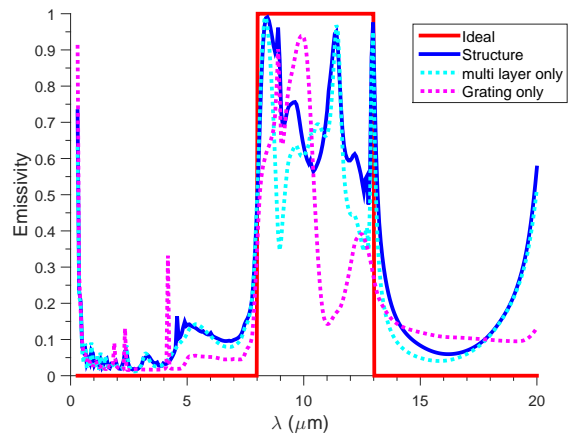}
 \caption{Emissivity at normal incidence as a function of wavelength : ideal (red line), optimized structure (blue line), multi-layer only (cyan) and grating alone with silver (magenta).}
 \label{decomp}
 \end{figure}

To illustrate our optimized structure, we separate in Fig.~\ref{decomp} the input of each component: ideal emissivity, optimized structure, grating alone and multi-layer.

\bigskip

The three emission peaks coming from the multi-layer structure are still found in the optimized structure emissivity. The peak at 9 $\mu$m from the grating is still there too. The second peak of the grating around 10 $\mu$m permits to keep the structure emissivity higher than with only the multi-layer. Finally the last contribution of the grating around 12 $\mu$m coupled with the multi-layer contribution produced a higher emissivity at this wavelength for the optimized structure. It is one example where the coupling between the multi-layer and the grating enhances the grating emissivity peaks. It could come either from the SPP excitation by the grating coupled to the multi-layer or from the grating acting as an additional layer to the multi-layer and enhancing its emissivity.

\subsection{Simplified structure for fabrication}
\begin{figure}[hbtp]
 \centering
 \includegraphics[width=7cm]{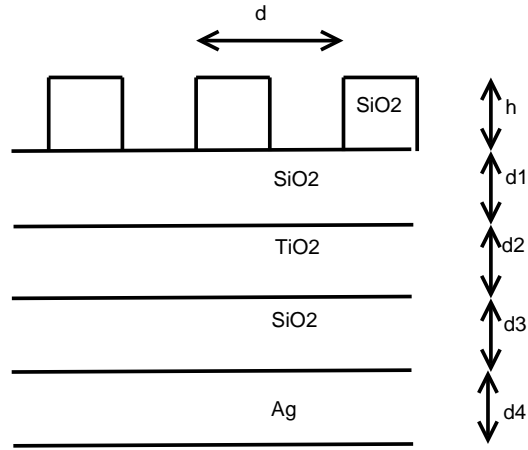}
 \caption{Simplified radiative cooler structure.}
 \label{exp}
 \end{figure}

\begin{figure}[hbtp]
 \centering
 \includegraphics[width=8cm]{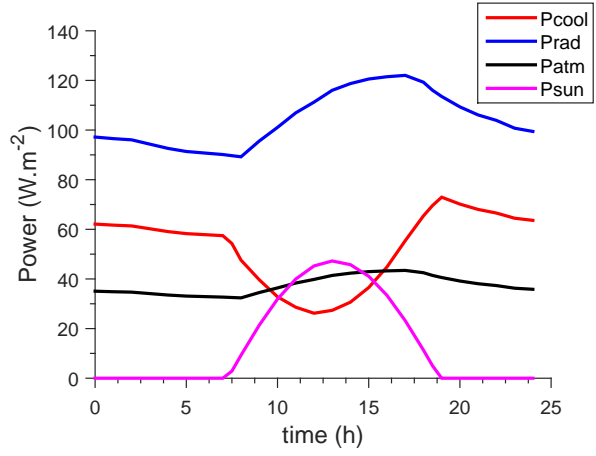}
 \caption{Radiative cooling power density (in red) of the simplified structure (with the average emissivity of the structure between TM and TE polarizations) as a function of time regarding to the temperature of a March day in Poitiers and its contributions $P_{rad}$ (in blue), $P_{atm}$ (in black) and $P_{sun}$ (in magenta).}
 \label{Pcool_E_8}
 \end{figure}

To simplify the fabrication, we next show that it is possible to have a good radiative cooling with a simplified structure composed of few layers. The design of this structure is shown in Fig.~\ref{exp}. Here, we complete the emission peak of the grating with the emission of an "optimized" Fabry-Perot cavity : a layer of TiO$_{2}$, located between two layers of SiO$_{2}$. 
In the same conditions as previously, we plot on Fig.~\ref{Pcool_E_8} the radiative cooling power density (in red) during the day for the average emissivity between TE and TM polarizations. We still have the maximum cooling power around 75 W.m$^{-2}$ at 7pm and the radiative cooling power density is still better at nighttime. Around noon, we attain a $P_{cool}$ of 30 W.m$^{-2}$. The performance of this simplified structure, a little bit under the performance of the precedent structure, is still very good for the considered external temperatures.

\bigskip

\begin{figure}[hbtp]
 \centering
 \includegraphics[width=8cm]{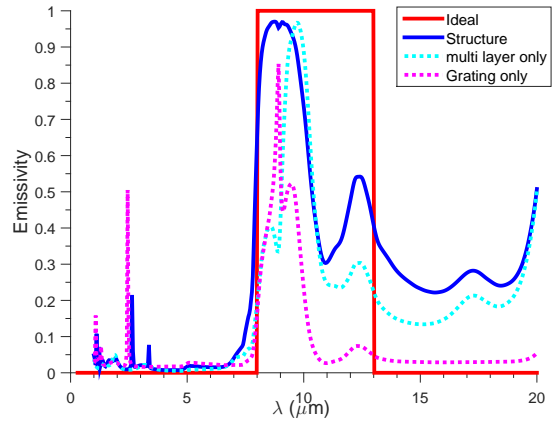}
 \caption{Emissivity at normal incidence as a function of wavelength : ideal (red line), optimized structure (blue line), multi-layer only (cyan) and grating alone with silver (magenta). }
 \label{decomp_simple}
 \end{figure}

As previously, we separate the input of each element in Fig.~\ref{decomp_simple}. The grating has an emission different from the other optimized structure because we changed its parameters $d$, $h$ and $F$. With its emission peak around 9 $\mu$m, the grating contributes greatly to the emission between 8 and 11 $\mu$m. Then the TiO$_{2}$ cavity produces a peak at 10 $\mu$m and a feeble one around 12 $\mu$m. Finally, it is then the combination of the multi-layer and the grating effects, which produces the great emission between 8 and 11 $\mu$m and increases the emissivity of the 12 $\mu$m-peak. This combination along with the silver layer also permits to reduce the emissivity in the solar spectrum.
This proves that the grating can, in addition to its own emission peaks, exhibit or enhance emissivity from the multi-layer structure below.

\bigskip

\section{Conclusion and perspectives}
We have shown that it is possible to design structures, which are very emissive in the atmospheric transparency window and very reflective in the solar spectrum, by associating thin film stacks with gratings. The coupling between polar material gratings and multi-layer structures creates additional emission peaks in the atmospheric transparency window and reduces emission elsewhere. Furthermore, the presence of the grating above the multi-layer can produce an emissivity greater than the emissivity made by the grating or multi-layer alone. This phenomenon could be a consequence of the SPP excitation in the multi-layer accentuated by the grating or the grating could play the role of an upper layer enhancing the emissivity of the multi-layer.

A simplified structure which produces a good radiative cooling power was also obtained. It will permit facilitating the experimental verification and as such validate the theoretical results. 

\bigskip

For this study, we worked only with 1D lamellar gratings as a preliminary work. 1D lamellar gratings can produce emission peaks for TE polarization and for azimutal angles different from zero. Nevertheless there is generally a distinction between the emission produced by both polarizations and 1D lamellar gratings produce emission peaks at best with a maximum of 50 \% for non polarized light and all emission angles. On the contrary, 2D gratings potentially produce emission peaks with a maximum of 100 \% for a non polarized light and all emission angles. It could improve further the radiative cooling power generated by our multi-layer + gratings structures. It could also help for practical applications in natural light as it could permit to be free of all polarization or azimuthal angle considerations. Thus a combination of a 2D grating with the simplified structure could be a good consideration for prototype fabrication and experimental tests will be done in this way soon.

\section*{Funding}
\textit{This work was partially funded by the French Government program ''Investissements d'Avenir'' (LABEX INTERACTIFS, reference ANR-11-LABX-0017-01) and ANR RADCOOL (reference ANR-17-CE06-0002-01).}


\begin{thebibliography}{99}

\bibitem{catalanotti_radiative_1975}
S.~Catalanotti, V.~Cuomo, G.~Piro, D.~Ruggi, V.~Silvestrini, and G.~Troise, ``The radiative cooling of selective surfaces,'' Sol. Energy
  {\bfseries 17}(2), 83--89 (1975).

\bibitem{bartoli_nocturnal_1977}
B.~Bartoli, S.~Catalanotti, B.~Coluzzi, V.~Cuomo, V.~Silvestrini, and
  G.~Troise,  ``Nocturnal and diurnal performances of selective
  radiators,'' Appl. Energy {\bfseries 3}(4), 267--286 (1977).

\bibitem{harrison_radiative_1978}
A.~W. Harrison and M.~R. Walton,  ``Radiative cooling of {TiO}2 white
  paint,'' Sol. Energy {\bfseries 20}(2), 185--188 (1978).

\bibitem{granqvist_surfaces_1980}
C.~G. Granqvist and A.~Hjortsberg,  ``Surfaces for radiative cooling:
  {Silicon} monoxide films on aluminum,'' \apl {\bfseries 36}(2),
  139--141 (1980).

\bibitem{granqvist_radiative_1981}
C.~G. Granqvist and A.~Hjortsberg,  ``Radiative cooling to low
  temperatures: general considerations and application to selectively
  emitting {SiO} films,'' J. Appl. Phys. {\bfseries 52}(6), 4205--4220
  (1981).

\bibitem{granqvist_radiative_1982}
C.~G. Granqvist, A.~Hjortsberg, and T.~S. Eriksson,  ``Radiative cooling
  to low temperatures with selectivity {IR}-emitting surfaces,'' Thin Solid
  Films {\bfseries 90}(2), 187--190 (1982).

\bibitem{berdahl_radiative_1984}
P.~Berdahl,  ``Radiative cooling with {MgO} and/or {LiF} layers,'' \ao {\bfseries 23}(3), 370--372 (1984).

\bibitem{berdahl_thermal_1983}
P.~Berdahl, M.~Martin, and F.~Sakkal,  ``Thermal performance of radiative
  cooling panels,'' Int. J. Heat Mass {\bfseries 26}(6),
  871--880 (1983).

\bibitem{orel_radiative_1993}
B.~Orel, M.~K. Gunde, and A.~Krainer,  ``Radiative cooling efficiency of
  white pigmented paints,'' Sol. Energy {\bfseries 50}(6), 477--482 (1993).

\bibitem{smith_amplified_2009}
G.~B. Smith,  ``Amplified radiative cooling via optimised combinations of
  aperture geometry and spectral emittance profiles of surfaces and the
  atmosphere,'' Sol. Energy Mater. Solar Cells {\bfseries 93}(9), 1696--1701
  (2009).

\bibitem{gentle_radiative_2010}
A.~R. Gentle and G.~B. Smith,  ``Radiative heat pumping from the
  earth using surface phonon resonant nanoparticles,'' Nano Lett.
  {\bfseries 10}(2), 373--379 (2010).

\bibitem{suryawanshi_radiative_2009}
C.~N. Suryawanshi and C.-T. Lin,  ``Radiative cooling: lattice quantization and surface emissivity in thin coatings,'' ACS Appl.
  Mater. Interfaces {\bfseries 1}(6), 1334--1338 (2009).

\bibitem{rephaeli_ultrabroadband_2013}
E.~Rephaeli, A.~Raman, and S.~Fan,  ``Ultrabroadband photonic structures to achieve high-performance daytime radiative cooling,'' Nano Lett. {\bfseries 13}(4), 1457--1461 (2013).

\bibitem{nilsson_solar_1992}
T.~M.~J. Nilsson, G.~A. Niklasson, and C.~G. Granqvist,  ``A solar
  reflecting material for radiative cooling applications: {ZnS} pigmented
  polyethylene,'' Sol. Energy Mater. Sol. Cells {\bfseries 28}(2), 175--193
  (1992).

\bibitem{nilsson_radiative_1995}
T.~M.~J. Nilsson and G.~A. Niklasson,  ``Radiative cooling during the
  day: simulations and experiments on pigmented polyethylene cover foils,''
  Sol. Energy Mater. Sol. Cells {\bfseries 37}(1), 93--118 (1995).

\bibitem{hossain_metamaterial_2015}
M.~M. Hossain, B.~Jia, and M.~Gu,  ``A metamaterial emitter for highly efficient radiative cooling,'' Advanced Optical Materials
  {\bfseries 3}(8), 1047--1051 (2015).

\bibitem{raman_passive_2014}
A.~P. Raman, M.~A. Anoma, L.~Zhu, E.~Rephaeli, and S.~Fan,  ``Passive
  radiative cooling below ambient air temperature under direct sunlight,''
  \nat {\bfseries 515}(7528), 540--544 (2014).

\bibitem{zhai_scalable-manufactured_2017}
Y.~Zhai, Y.~Ma, S.~N. David, D.~Zhao, R.~Lou, G.~Tan, R.~Yang, and X.~Yin,
   ``Scalable-manufactured randomized glass-polymer hybrid metamaterial
  for daytime radiative cooling,'' Science {\bfseries 355}(6329), 1062--1066 (2017).

\bibitem{zhu_radiative_2014}
L.~Zhu, A.~Raman, K.~X. Wang, M.~A. Anoma, and S.~Fan,  ``Radiative
  cooling of solar cells,'' Optica {\bfseries 1}(1), 32--38 (2014).

\bibitem{zhu_radiative_2015}
L.~Zhu, A.~P. Raman, and S.~Fan,  ``Radiative cooling of solar absorbers
  using a visibly transparent photonic crystal thermal blackbody,'' Proceedings
  of the National Academy of Sciences {\bfseries 112}(40), 12282--12287 (2015).

\bibitem{safi_improving_2015}
T.~S. Safi and J.~N. Munday,  ``Improving photovoltaic performance
  through radiative cooling in both terrestrial and extraterrestrial
  environments,'' \opex {\bfseries 23}(19), A1120--1128 (2015).

\bibitem{wu_solar_2015}
S.-H. Wu and M.~L. Povinelli,  ``Solar heating of {GaAs} nanowire solar
  cells,'' \opex {\bfseries 23}(24), A1363--1372 (2015).

\bibitem{zhou_radiative_2016}
Z.~Zhou, X.~Sun, and P.~Bermel,  ``Radiative cooling for
  thermophotovoltaic systems,'' Proc. SPIE {\bfseries 9973}, 997308 (2016).

\bibitem{huang_nanoparticle_2017}
Z.~Huang and X.~Ruan,  ``Nanoparticle embedded double-layer coating for
  daytime radiative cooling,'' Int. J. Heat Mass {\bfseries 104}, 890--896 (2017).

\bibitem{kecebas_passive_2017}
M.~A. Kecebas, M.~P. Menguc, A.~Kosar, and K.~Sendur,  ``Passive
  radiative cooling design with broadband optical thin-film filters,'' J. Quant. Spectro. Rad. Trans. {\bfseries 198}, 179--186 (2017).

\bibitem{kou_daytime_2017}
J.-l. Kou, Z.~Jurado, Z.~Chen, S.~Fan, and A.~J. Minnich,  ``Daytime radiative cooling using near-black infrared emitters,'' ACS
  Photonics {\bfseries 4}(3), 626--630 (2017).

\bibitem{greffet_coherent_2002}
J.-J. Greffet, R.~Carminati, K.~Joulain, J.-P. Mulet, S.~Mainguy, and Y.~Chen,
   ``Coherent emission of light by thermal sources,'' \nat {\bfseries 416}(6876),
  61--64 (2002).

\bibitem{marquier_coherent_2004}
F.~Marquier, K.~Joulain, J.-P. Mulet, R.~Carminati, J.-J. Greffet, and Y.~Chen,
   ``Coherent spontaneous emission of light by thermal sources,'' \prb {\bfseries 69}(15), 155412 (2004).

\bibitem{dahan_space-variant_2005}
N.~Dahan, A.~Niv, G.~Biener, V.~Kleiner, and E.~Hasman,  ``Space-variant
  polarization manipulation of a thermal emission by a {SiO}2 subwavelength
  grating supporting surface phonon-polaritons,'' \apl {\bfseries 86}(19), 191102 (2005).

\bibitem{arnold_coherent_2012}
C.~Arnold, F.~Marquier, M.~Garin, F.~Pardo, S.~Collin, N.~Bardou, J.-L.
  Pelouard, and J.-J. Greffet,  ``Coherent thermal infrared emission by
  two-dimensional silicon carbide gratings,'' \prb {\bfseries 86}(3),
  035316 (2012).

\bibitem{drevillon_far_2011}
J.~Drevillon, K.~Joulain, P.~Ben-Abdallah, and E.~Nefzaoui,  ``Far field
  coherent thermal emission from a bilayer structure,'' J. Appl.Phys. {\bfseries 109}(3), 034315 (2011).

\bibitem{nefzaoui_selective_2012}
E.~Nefzaoui, J.~Drevillon, and K.~Joulain,  ``Selective emitters design
  and optimization for thermophotovoltaic applications,'' J. Appl. Phys. {\bfseries 111}(8), 084316 (2012).

\bibitem{marquier_anisotropic_2006}
F.~Marquier, M.~Laroche, R.~Carminati, and J.-J. Greffet,  ``Anisotropic polarized emission of a doped silicon lamellar grating,'' J. Heat Trans. {\bfseries 129}(1), 11--16 (2006).

\bibitem{marquier_degree_2008}
F.~Marquier, C.~Arnold, M.~Laroche, J.~J. Greffet, and Y.~Chen,  ``Degree
  of polarization of thermal light emitted by gratings supporting surface
  waves,'' \opex {\bfseries 16}(8), 5305 (2008).

\bibitem{herve_temperature_2016}
A.~Herv\'{e}, J.~Dr\'{e}villon, Y.~Ezzahri, K.~Joulain, D.~De~Sousa~Meneses, and
  J.-P. Hugonin,  ``Temperature dependence of a microstructured {SiC}
  coherent thermal source,'' J. Quant. Spectro. Rad. Trans. {\bfseries 180}, 29--38 (2016).

\bibitem{kennedy_particle_1995}
J.~Kennedy and R.~Eberhart,  ``Particle swarm optimization,'' Proc. Int.
  Conf. Neural Networks {\bfseries 4}, 1942--1948 (1995).

\bibitem{clerc_optimisation_2004}
M.~Clerc,  {\itshape L'optimisation par essaim particulaires}, Hermes {Science}
  {Publications} (2004).

\bibitem{chen_radiative_2016}
Z.~Chen, L.~Zhu, A.~Raman, and S. Fan,  ``Radiative cooling to deep sub-freezing temperatures through a 24-h day-night cycle,'' Nature Communications
  Transfer {\bfseries 7}, 13729 (2016).
\end{thebibliography}
\end{document}